\documentclass[preprint, showpacs,preprintnumbers,amsmath,amssymb,showkeys]{revtex4}
\usepackage{graphicx}
\usepackage{dcolumn}
\usepackage{bm}

\def\be{\begin{equation}}
\def\ee{\end{equation}}
\def\bea{\begin{eqnarray}}
\def\eea{\end{eqnarray}}

\begin{document}

\title{Skyrme model and Isospin Chemical Potential}
\author{M. Loewe}
\email{mloewe@fis.puc.cl} \affiliation{Facultad de F\'\i sica,
Pontificia Universidad Cat\'olica de Chile,\\ Casilla 306, Santiago
22, Chile.}
\author{S. Mendizabal}
\email{smendiza@fis.puc.cl} \affiliation{Facultad de F\'\i sica,
Pontificia Universidad Cat\'olica de Chile,\\ Casilla 306,
Santiago 22, Chile.}
\author{J.C. Rojas}
\email{jurojas@ucn.cl} \affiliation{Departamento de F\'{\i}sica,
Universidad Cat\'{o}lica del Norte,\\ Casilla 1280, Antofagasta,
Chile}

\begin{abstract}
We discuss the stability of the Skyrmion solution in the presence
of a finite isospin chemical potential $\mu$. Solving numerically
the mass of the Skyrmion as function of $\mu$, we find a critical
value $\mu_c=222.8$ MeV where the Skyrmion mass vanishes. We
compare the exact numerical treatment with an analytical
discussion based on a special shape for the profile of the
Skyrmion due to Atiyah and Manton. The extension of this ansatz
for finite $\mu$ works quite well for $\mu<121$ MeV. Then, for
small values of $\mu$, where the analytical approach is valid, we
consider the possibility of having an angular deformation for the
Skyrmionic profile, which is possible for finite values of $\mu$.
This is however, a small effect. Finally we introduce finite
temperature corrections, which strength the instability induced by
the chemical potential, finding the dependence of the critical
temperature on $\mu$.

\end{abstract}

\maketitle

\section{Introduction}

The skyrmion picture \cite{skyrme} has attracted the attention of
many people as
 a possible way for understanding the hadronic dynamics, as well as the hadronic
 phase structure. The behavior of hadrons in the presence of a media,
i.e. taking into account temperature and/or density effects, can
be analyzed according to this perspective. For example, in a
baryon-rich environment, it is found that coupling constants like
$g_A$ and $F_{\pi}$ become quenched \cite{rho}. In reference
\cite{kajantie,dey}, which extends the construction of Atiyah and
Manton in \cite{Manton}, the authors discussed the stability of
skyrmions under finite temperature conditions, showing the
existence of a critical temperature $T_c$. The skyrmion is not
longer stable for $T>T_c$.

\bigskip

A similar kind of discussion has been carried out in the frame of
the so called hybrid models \cite{falomir}. The construction of
Hadrons as a core, described trough a chiral bag, surrounded by
the tail of a skyrmion, has the same instability properties for a
certain critical temperature. This can be interpreted as the
occurrence of a deconfining phase transition.

\bigskip

Recently \cite{topo}, we have analyzed different topological
structures in field theory when a finite isospin chemical
potential ($\mu$) is taken into account. In particular, we were
able to find an upper bound $\mu_c$ for the isospin chemical
potential, such that the mass of the skyrmion vanishes.

\bigskip

In this letter, we improve our previous discussion \cite{topo},
showing that, in fact, the skyrmion develops an instability as
function of $\mu$, in the sense that the mass of the skyrmion,
$M=M(\mu)$ diminishes and vanishes at a certain critical value
$\mu_c$. We found $\mu_c$ through an exact numerical analysis of
the Skyrmion mass evolution, without referring to any particular
radial profile. In a second stage, we compare the numerical
results with an analytical treatment based on the Atiyah-Manton
ansatz \cite{Manton}. We would like to emphasize that this profile
is obtained from a construction based on $SU(2)$ instantons in
four dimensions, by computing Holonomies along lines parallel to
the time axis. In this sense, this profile is definitely more
fundamental than just an educated guess. The comparison between
the numerical analysis, and the analytical procedure works quite
well, in a wide range of $\mu$ up to a certain value of the order
of $\mu\approx 110 MeV$, where in the analytical approach it is
not longer possible to get a stable mass for the Skyrmion.

\bigskip

In addition, we propose an extension of the radial ansatz of the
skyrmion when $\mu \neq 0$, showing that a small angular
deformation occurs for the minimum energy configuration, for small
values of $\mu$. This is done in the frame of a natural extension
of Atiyah-Manton's ansatz. Finally we include temperature
corrections, finding how the critical temperature depends on
$\mu$.

\section{Isospin Chemical Potential and Skyrmion stability}

The Skyrme lagrangian is

\bea {\cal L} &=& \frac{F_{\pi}^2}{16} Tr\left[
\partial_{\mu}U \partial^{\mu}U^{\dagger}
\right] \nonumber \\ &&+ \frac{1}{32e^2}Tr \left[
(\partial_{\mu}U)U^{\dagger}, (\partial_{\nu}U)U^{\dagger}
\right]^2, \eea

\noindent where $F_{\pi}$ is the pion decay constant and $e$ is a
numerical parameter, following the notation used in
\cite{ad-witt}. The introduction of an isospin chemical potential
can be easily done by introducing a covariant derivative of the
form \cite{actor,weldon}

\be
\partial_{\nu} U \rightarrow  D_{\nu}U = \partial_{\nu}U -i \frac{\mu}{2} [\sigma^3,U] g_{\nu
0}. \ee

\noindent Defining $L_{\nu} \equiv (\partial_{\nu}U) U^{\dagger}$,
 the lagrangian density for finite $\mu$ becomes

\bea {\cal L}_{\mu} &=&
 \frac{F_{\pi}^2}{16} Tr \left\{
 \partial_{\nu}U \partial^{\nu}U^{\dagger} + \frac{\mu^2}{2}
\left[ \sigma_0 - U \sigma_3 U^{\dagger} \sigma_3 \right] \right\}
\nonumber \\
 &+& \frac{1}{32e^2} Tr\left\{ \left[L_{\rho},L_{\nu} \right]^2 -
\frac{\mu^2}{2} \left( [\varrho,L_{\nu}][\varrho,L^{\nu}] \right)
\right\}, \eea

\noindent where $\varrho = \sigma_3-U\sigma^3 U^{\dagger}$ and
$\sigma_0$ is the $2 \times 2$ unit matrix.

\bigskip

With a little algebra, we can split the lagrangian into a zero and
a finite chemical potential contributions

\bea {\cal L}_{\mu} &=& {\cal L}_{\mu=0} + \frac{F_{\pi}^2
\mu^2}{32} Tr \left[ \sigma_0 - U \sigma_3 U^{\dagger} \sigma_3
\right]
\nonumber\\
&& + \frac{\mu^2}{64 e^2} Tr  \left[ \varrho,L_{\nu} \right]^2.
\eea

\noindent ${\cal L}_{\mu=0}$ denotes the zero chemical potential
term.

\bigskip

The $U$ field matrix can be parameterized in the standard way

\be U = \exp \left( -i\xi \vec{\sigma}\cdot \hat{n}
\right)=\cos\xi -i(\vec{\sigma} \cdot \hat{n})\sin\xi, \ee

\noindent where $\vec{\sigma}$ is the sigma matrix vector and
$\hat{n}^2=1$. This ansatz has a ``Hedgehog" shape and the
following boundary conditions have to be satisfied \cite{ad-witt}

\bea &&\xi(\vec{r})=\xi(r), \; \; \hat{n}=\hat{r},
\nonumber \\
&&\xi(0)=\pi, \; \; \xi(\infty)=0. \eea

The mass of the Skyrmion, for static solutions, develops a
dependence on the Isospin Chemical potential as well as on the
temperature. For the remaining of this section, we will
concentrate on the $T=0$ scenario. If we define the
non-dimensional variable $\hat{r}=e F_{\pi} r$, the mass of the
Skyrmion will be given by

\be M_{\mu}=M_{\mu=0}- \frac{\mu^2}{4 e^3 F_{\pi}} I_2 -
\frac{\mu^2}{32 e^3 F_{\pi}} I_4, \label{mmu}\ee

\noindent where $M_{\mu=0}$ is the zero chemical potential
contribution

\bea M_{\mu=0} &=& \frac{F_{\pi}}{4 e} \left\{ 4 \pi
\int_0^{\infty}d\hat{r} \left[ \frac{\hat{r}^2}{2} \left(
\frac{d\xi_1}{d\hat{r}}\right)^2 + \sin^2(\xi_1)\right]\right.
\nonumber
\\  && + 4 \pi \int_0^{\infty}d\hat{r}
\frac{\sin^2(\xi_1)}{\hat{r}^2} \times \nonumber \\ & &  \left.
\left[ 4\hat{r}^2 \left(
\frac{d\xi_1}{d\hat{r}}\right)^2+2\sin^2(\xi_1)\right]
\right\},\eea

\noindent and $I_2$ and $I_4$ are the following non-dimensional
integrals

\bea I_2 &=& \int d^3\hat{r} \mathrm{Tr} \left[ \sigma_0
-U\sigma_3U^{\dagger}\sigma_3\right],
\nonumber \\
I_4 &=&  \int d^3\hat{r} \mathrm{Tr} \left[ \varrho,L_{\nu}
\right]^2. \label{Ies}\eea

We would like to emphasize that the chemical potential terms
contribute with opposite sign. This implies that the solution
becomes unstable above certain value of $\mu$.

\bigskip

Assuming a radial symmetric profile ($\xi=\xi(r)$), we can
evaluate directly the traces in (\ref{Ies}), getting

\bea \mathrm{Tr} \left[ \sigma_0 - U \sigma_3 U^{\dagger} \sigma_3
\right]
&=& 4 \sin^2\theta \sin^2\xi, \\
\mathrm{Tr} \left[ \varrho,L_{\nu} \right]^2 =&& 32 \sin^2\theta
\sin^2\xi \left( \frac{d\xi}{d \hat{r}} \right)^2\nonumber\\ &+&
4\left(\frac{32}{\hat{r}^2}\right) \sin^2\theta \sin^4\xi, \eea

\noindent leading us to the following integrals

\bea I_2 &=& \frac{4 \pi}{3} \int d\hat{r} \hat{r}^2 \sin^2 \xi,\\
I_4 &=& \frac{32 \pi}{3} \int d\hat{r} \hat{r}^2 \left[ \sin^2 \xi
\left( \frac{d\xi}{d\hat{r}}\right)^2+ \frac{4}{\hat{r}^2} \sin^4
\xi \right]. \label{integrals}\eea

\begin{figure}
\includegraphics[angle=0,width=0.42\textwidth]{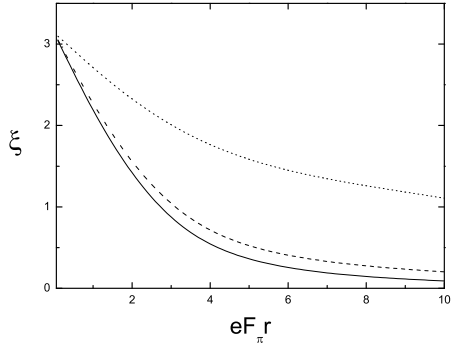}
\caption{\label{perfilvsradio} Numerical solution for the profile
$\xi(r)$ for different values of $\mu$. ($\mu=12.9$ (MeV): solid
line; $\mu=25.8$ (MeV): dashed line; $\mu=38.7$ (MeV): dotted
line.)}
\end{figure}

The variational equation for the profile according to (\ref{mmu})
is
\begin{widetext}
\begin{eqnarray}
&&\left(\frac{1}{4}\hat{r}^2+2\sin^{2}{\xi}\right)\frac{d^2\xi}{d\hat{r}^2}+\frac{1}{2}\hat{r}\frac{d\xi}{d\hat{r}}+
\sin{2\xi}\left(\frac{d\xi}{d\hat{r}}\right)^{2}-\frac{1}{4}\sin{2\xi}-\frac{\sin^{2}
{\xi}\sin{2\xi}}{\hat{r}^{2}}\nonumber\\&&
-\frac{\hat{\mu}^{2}\hat{r}^{2}\sin^{2}{\xi}}{3}\left(\frac{1}{2}\frac{d^2\xi}{d\hat{r}^2}
+\frac{1}{2\sin{\xi}}\left(\frac{d\xi}{d\hat{r}}\right)^{2}+
\frac{1}{\hat{r}}\frac{d\xi}{d\hat{r}}-\frac{1}{4}\frac{\sin{2\xi}}{\sin^2{\xi}}-\frac{2\sin{2\xi}}{\hat{r}}\right)=0,
\label{ecmovi}
\end{eqnarray}

\end{widetext}

\noindent where $\hat{\mu}=\mu/(eF_{\pi})$. Equation
(\ref{ecmovi}) can be solved numerically for the profile $\xi(r)$,
for different values of $\mu$ (Figure \ref{perfilvsradio}). Notice
that the radio extension of the Skyrmion grows as a function of
$\mu$.

\bigskip

\begin{figure}
\includegraphics[angle=0,width=0.45\textwidth]{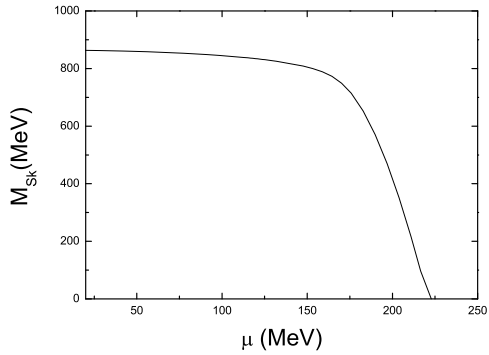}
\caption{\label{numericomasa} The numerical solution for the
Skyrmion mass as funtion of $\mu$.}
\end{figure}

In order to obtain the mass of the Skyrmion, the profile has to be
inserted, numerically, in equation (\ref{mmu}). Figure
\ref{numericomasa} shows the chemical potential dependence of the
mass. The point where the mass vanishes corresponds to the
critical value $\mu_c=222.8$ MeV. This value does not depend on a
model for the shape of $\xi(r)$. It is a fundamental result
associated to the Skyrmion picture.

\bigskip

The same analysis can be done using a specific profile for
$\xi(r)$. Here we will use the ansatz introduced by Atiyah and
Manton in \cite{Manton}, which depends on a free parameter
$\lambda$:

\be \xi_{\lambda}(r)= \pi \left[ 1-\frac{r}{\sqrt{r^2+\lambda^2}}
\right]. \label{xi0}\ee

The variation with respect to $\lambda$ enables us to find the
lower mass of the Skyrmion. This approach has proved to fit very
accurately the numerical value of the mass of the Skyrmion in the
$\mu=0$.

\bigskip

The function $\xi$ can be scaled

\bea \xi_{\lambda}(r)&=& \xi_{1}(r/\lambda), \\
\frac{d\xi_{\lambda}(r)}{dr} &=&
\frac{1}{\lambda}\frac{d\xi_{1}(r/\lambda)}{d(r/\lambda)}, \eea

\noindent in order to factorize $\lambda$ from the integrals in
(\ref{integrals}). Using $y=r/\lambda=\hat{r}/\tilde{\lambda}$,
with $\tilde{\lambda}=eF_{\pi}\lambda$, we obtain

\bea I_2 &=& \tilde{\lambda}^3 \left(\frac{4 \pi}{3} \right)
\int_0^{\infty} dy y^2 \sin^2 (\xi_1(y)/2)\nonumber \\ &=&
1.927 \left(\frac{32 \pi}{3} \right) \tilde{\lambda}^3,\\
I_4 &=& \tilde{\lambda} \left( \frac{4 \pi}{3}
\right)\int_0^{\infty} dy y^2 \times \nonumber \\ && \left[ \sin^2
(\xi_1(y)/2) \left( \frac{d\xi_1}{dy}\right)^2 +
\frac{4}{y^2} \sin^4 (\xi_1(y)/2) \right], \nonumber \\
&=&  3.728 \left( \frac{64 \pi}{3} \right) \tilde{\lambda}. \eea

Defining the dimensionless quantities $\tilde{M}_{\mu}\equiv
M_{\mu}e/F_{\pi}$ and $\tilde{\mu}\equiv \mu/(eF_{\pi})$, we find
the Skyrmion mass as a function of the free parameter $\lambda$

\bea \tilde{M}_{\mu} =&&  \left( 4\pi \right) \left[ 0.504
\tilde{\lambda}+4.254 \frac{1}{\tilde{\lambda}} -0.101
\tilde{\mu}^2
\tilde{\lambda}^3\right.\nonumber\\
&-&0.311\tilde{\mu}^2\tilde{\lambda} \left.\right] \eea

\begin{figure}
\includegraphics[angle=0,width=0.45\textwidth]{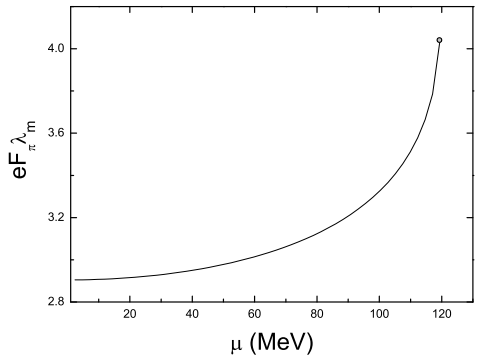}
\caption{\label{fig1} The evolution of the
$\tilde{\lambda}_m/F_{\pi}$ parameter as function of $\mu/F_{\pi}$
(The dot denotes the critical point).}
\end{figure}

The $\tilde{\lambda}_m$ which minimizes the expression takes the
value

\bea \tilde{\lambda}^2_m =&& 0.523\,{\frac
{1}{{\tilde{\mu}}^{2}}}-
0.322\nonumber\\
 &-& \frac{1.038}{\tilde{\mu}^2}\sqrt {{ 0.254- 8.511\,{\tilde{\mu}}^{ 2}+ 0.097\,\tilde{\mu}}^{4}},
 \eea

\noindent where we have used the dimensionless parameter
$\tilde{\lambda}$ and $e=5.45$ according to \cite{ad-witt}. In
Figure \ref{fig1} we show the dependence of $\tilde{\lambda}_m$ on
$\mu$. In the limit $\mu\rightarrow0$, we recover the result
$\tilde{\lambda}_m=2.904$ \cite{Manton}. Notice that we have an
upper bound for $\tilde{\mu}=\tilde{\mu}_c \rightarrow 0.173$,
where $\tilde{\lambda}_m$ takes the value
$\tilde{\lambda}_m=4.138$. For bigger values of $\tilde{\mu}$,
$\tilde{\lambda}_m$ becomes imaginary, implying, that the Skyrmion
is no longer stable in the regime $\tilde{\mu}>\tilde{\mu}_{c}$.
The mass of the Skyrmion, as function of $\mu$, shows a small
decrease, about $8\%$.

\begin{figure}
\includegraphics[angle=0,width=0.45\textwidth]{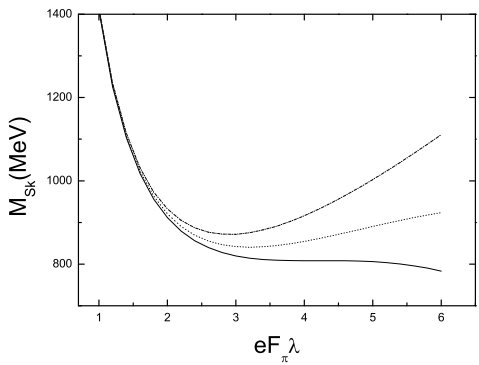}
\caption{\label{fig3} The Skyrmion mass as function of $\lambda$
for three values of $\mu$ ($\hat{\mu}=1.29$(MeV): dot-dashed line;
$\hat{\mu}=64.5$ (MeV): dotted line; $\hat{\mu}=118.68$ (MeV):
continuous line ).}
\end{figure}

\bigskip

In Figure \ref{fig3} we show how the mass dependence on $\lambda$
evolves for three different values of the chemical potential. Note
that for the $\tilde{\mu}=0.926$, the minimum disappears and we
have an inflection point signalizing the end of the stable
Skyrmion phase.\\

\bigskip

\begin{figure}
\includegraphics[angle=0,width=0.45\textwidth]{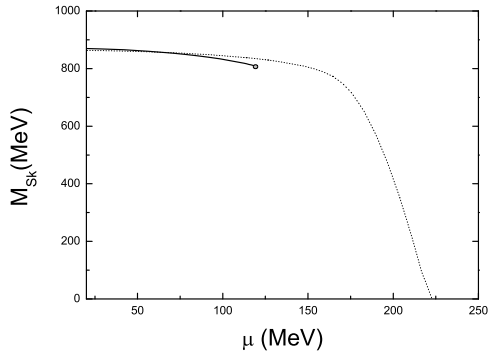}
\caption{\label{fig6} Comparison between the exact numerical
solution (dashed line) and the Atiyah-Manton's ansatz (continous
line) for the mass of the Skyrmion.}
\end{figure}

 In figure \ref{fig6} we compare the numerical
solution for the mass with our analytical result. We see that in a
wide region, up to $\mu\approx 100$ MeV, the discrepancy between
both approaches is less than $2\%$, showing that the ansatz is a
very good approximation in this region.

Notice that the critical value $\mu_c$ where the mass of the
Skyrmion vanishes, is much lower than the one we obtained in our
previous article \cite{topo}, using the Atiyah-Manton's ansatz
with a fixed value for $\lambda$.

\section{Anisotropic Skyrmion Ansatz}

If $\mu \neq 0$, we may expect the occurrence of anisotropic
stable skyrmions, i.e. solutions where the profile $\xi$ becomes
also dependent on the azimuthal angle $\theta$
$(\xi=\xi(r,\theta))$. The discussion that follows is based on an
appropriate extension of Atiyah-Manton's construction in a region
of $\mu$, where we had almost a coincidence between the numerical
and analytical solution. A possible simple ansatz which decouples
the angular dependence in the integrals (\ref{mmu}), is the
following

\be \xi_{\lambda}(r,\theta)= 2 \pi \left[1-\frac{r
f(\theta)}{\sqrt{r^2 f^2(\theta)+\lambda^2}} \right],\ee

\noindent where the function $f(\theta)$ represents a small
deformation with respect to the radial configuration. A natural
candidate for $f(\theta)$ is to take the first term in a Fourier
expansion:

\be f(\theta)=1+a \sin \theta+b \cos \theta.\ee

\noindent We found that the cosine term does not contribute to
diminish the Skyrmion mass, so we can take $b=0$. In this way, the
energy equation is modified, due to the angular terms, according
to

\bea M_{\mu,a} &=& \frac{F_{\pi}\hat{\lambda}}{8 e}  (\tilde{I}_1
+\tilde{I}_2) + \frac{F_{\pi}}{32 e \hat{\lambda}} (\tilde{I}_3
+\tilde{I}_4) \nonumber \\ &-& \frac{\mu^2 \hat{\lambda}^3}{4 e^3
F_{\pi}}\tilde{I}_5- \frac{\mu^2 \hat{\lambda}}{32 e^3
F_{\pi}}\tilde{I}_{6}.\label{mmua}\eea

\noindent In the previous expression, we have introduced the
following set of integrals.

\bea \tilde{I}_1 &=& 2 \pi \int dr r^2 d\theta \sin\theta \;
\Big{[}(\partial_r \xi_1)^2+\frac{1}{r^2}
(\partial_{\theta}\xi_1)^2
\Big{]} \nonumber \\
\tilde{I}_2 &=& 2 \pi \int dr r^2 d\theta \sin\theta \;
\Big{[}1-\cos\xi_1\Big{]} (\partial_i\hat{n})^2 \nonumber \\
\tilde{I}_3 &=& \pi \int dr r^2 d\theta \sin\theta
\Big{[}1-\cos\xi_1\Big{]} \nonumber \\ && \times (\partial_i\xi_1
\partial_j\hat{n}-\partial_j\xi_1 \partial_i\hat{n})^2 \nonumber \\ \tilde{I}_4 &=&
2 \pi \int dr r^2 d\theta \sin\theta \; (1-\cos\xi_1)^2
(\partial_i\hat{n} \times \partial_j\hat{n})^2 \nonumber \\
\tilde{I}_5 &=& 8 \pi \int dr r^2 d\theta \sin^3\theta \; \theta
\sin^2(\xi_1/2) \nonumber \\
\tilde{I}_{6} &=& 2 \pi \int dr r^2 d\theta \sin\theta \Big{\{}
8\sin^2\theta \sin^2(\xi_1/2)  \nonumber \\ && \times
\Big{[}(\partial_r \xi_1)^2+\frac{1}{r^2}
(\partial_{\theta}\xi_1)^2 \Big{]}\nonumber \\ && +
\frac{32}{r^2}\sin^2\theta \sin^4(\xi_1/2) \Big{\}}.\eea

\begin{figure}
\includegraphics[angle=0,width=0.45\textwidth]{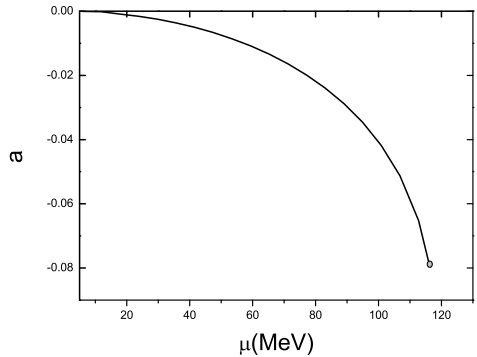}
\caption{\label{fig4} The dependence of the deformation parameter
$a$ as function of $\mu$.}
\end{figure}

In Fig \ref{fig4} we have plotted the dependence of the
deformation parameter $a(\mu)$, showing that there is a small
shift in the shape of the Skyrmions. The criteria for determining
the dependence of $a(\mu)$ is to minimize equation (\ref{mmua}).
Such deformation is possible due to the special character of the
Skyrmionic solution which is a topological object that entangles
the isospin with the spatial
 degrees of freedom and therefore, a finite isospin chemical potential has a nontrivial
 consequences.

\section{Finite temperature effects}

\begin{figure}[hb]
\includegraphics[angle=0,width=0.45\textwidth]{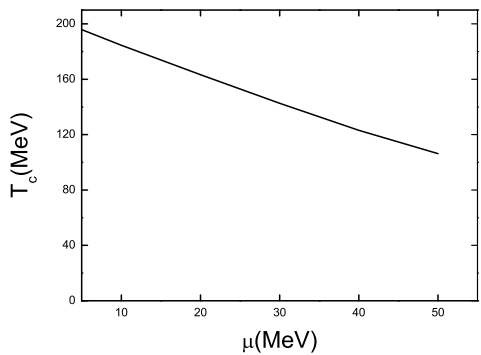}
\caption{\label{tcrit} The dependence of the critical temperature
 on $\mu$.}
\end{figure}

Finally, we will include temperature effects in our discussion,
following closely \cite{kajantie}, using their thermal profile for
the Skyrmion

\be \xi_{\lambda}^{\kappa}(r) = 2 \pi
\left[1-\frac{r+\frac{1}{2}\lambda^2(\kappa\coth(\kappa
r)-1/r)}{\sqrt{r^2+\frac{1}{4}\kappa^2 \lambda^4+\kappa r
\lambda^2 \coth(\kappa r)}} \right], \label{xitemp}\ee

\bigskip

\noindent where $\kappa=2 \pi T$. In the limit
$\kappa\rightarrow0$, we recover equation (\ref{xi0}). Such
profile was derived through an extension to finite temperature of
the instanton construction by Atiyah and Manton \cite{Manton}.

\bigskip

In this part of the analysis we will not take into account
possible angular deformations of the Skyrmion, since temperature
does not affect the shape of the Skyrmion. Also the onset of
instability due to chemical potential effects will occur for lower
values of $\mu$ and therefore we can disregard it.

\bigskip

The procedure we follow is to insert the thermal Skyrmion ansatz
(\ref{xitemp}) in equation (\ref{mmu}). Then we minimize with
respect to the parameter $\lambda$, getting for a certain $\mu$,
the critical temperature $T_c$ where the solution cannot attain a
minimum, becoming unstable.

\bigskip

In figure (\ref{tcrit}) we show how $T_c$ depends un $\mu$. It is
clear that $T_c$ diminishes with $\mu$. Both parameters,
temperature and chemical potential, cooperate in order to increase
the occurrence of instabilities in the Skyrmion solution.\\

\bigskip

In this letter we have discussed the stability of the Skyrmion
configuration in the presence of finite isospin chemical potential
($\mu$), showing the existence of a critical value $\mu_c$ where
the mass vanishes. Since the idea behind the Skyrmion approach is
to have an effective representation of baryons, the existence of
such phase transition is quite interesting. There are in the
literature other discussions of phase transitions induced by
density and temperature. For example, in the frame of chiral
Lagrangians, a detailed analysis of the phases of pion systems
\cite{villa} has been carried out, showing the existence of phase
transitions with the same kind of behavior for the masses that we
have presented here, in a complete different approach.

\section*{ACKNOWLEDGMENTS}

The authors would like to thank financial support from
 FONDECYT under grant 1051067.


\begin{thebibliography}{99}

\bibitem{skyrme} T.H.R. Skyrme, Proc.Roy.Soc.Lond.\textbf{A}260 (1961)127.

\bibitem{rho}  M. Rho, Phys.Rev.Lett.54 (1985) 767.

\bibitem{kajantie} K.J. Eskola and K. Kajantie, Z.f$\ddot{u}$r
Phys.\textbf{C}44 347 (1989) 347.

\bibitem{dey} J. Dey and J.M. Eisenberg, Phys.Lett.\textbf{B}334 (1994) 290.

\bibitem{Manton} M.F. Atiyah and N.S. Manton, Phys.Lett.\text{B}
222 (1989) 438.

\bibitem{falomir} H. Falomir, M. Loewe, J.C. Rojas,  Phys.Lett.\textbf{B}300
(1993) 278.

\bibitem{topo} M. Loewe, S. Mendizabal, J.C. Rojas, Phys.Lett.\textbf{B}609
(2005) 437.


\bibitem{ad-witt} G.S. Adkins, C.R. Nappi And E. Witten, Nucl.
Phys.\textbf{B}228 (1983) 552.

\bibitem{actor} A. Actor, Phys.Lett.\textbf{B}157 (1985) 53.

\bibitem{weldon} H.Arthur Weldon, Phys.Rev.\textbf{D}26 (1982)
1394.

\bibitem{villa} M. Loewe And C. villavicencio,
Phys.Rev.\textbf{B}71 (2005) 094001 and references therein.




\end{thebibliography}
\end{document}